\documentclass[pra,aps,amsfonts,amsmath,superscriptaddress,twocolumn,nopacs,floatfix]{revtex4}
\usepackage{graphicx} 
\usepackage{float} 
\usepackage{epsfig} 
\usepackage{subfigure}
\usepackage{rotating}

\begin{document}

\title{Cluster-liquid transition in finite saturated fermionic systems}
\author{J.-P. Ebran}
\affiliation{CEA,DAM,DIF, F-91297 Arpajon, France}
\author{E. Khan}
\affiliation{Institut de Physique Nucl\'eaire, Universit\'e Paris-Sud, IN2P3-CNRS, F-91406 Orsay Cedex, France}
\author{T. Nik\v si\' c}
\author{D. Vretenar}
\affiliation{Physics Department, Faculty of Science, University of
Zagreb, 10000 Zagreb, Croatia}

\begin{abstract}
The role of saturation for cluster formation in atomic nuclei is analyzed 
by considering three length-scale ratios, 
and performing deformation-constrained self-consistent mean-field 
calculations. The effect of clusterization in deformed light systems is 
related to the saturation property of the inter-nucleon interaction.
The formation of clusters at low nucleon density is illustrated 
by expanding the radius of $^{16}$O in a constrained calculation. 
A phase diagram shows that
the formation of clusters can be interpreted as a hybrid state 
between the crystal and the liquid
phases. In the hybrid cluster phase the confining potential
attenuates the delocalization generated by the effective
nuclear interaction. 

\end{abstract}

%\pacs{21.10.Re, 21.65.-f, 21.60.Jz}

\date{\today}

\maketitle

%\vspace{2cm}
When temperature decreases and density
increases, a system of particles interacting through a short-range
force undergoes a transition from a classical gaseous state to a
liquid one. Lowering the temperature further, in most cases 
a first-order phase transition from the liquid state to a solid state 
occurs. Increasing the density, that is, adding constituents between 
the nodes of the microscopic crystal structure, their wave functions 
start to overlap and clusters can be formed. With a further increase 
of the density the system becomes more homogeneous, finally reaching the
quantum liquid state \cite{pines}. Quantum effects become important 
when the typical dispersion of the constituent particles, that is, the 
thermal de Broglie wavelength of a particle 
\begin{equation}\label{eq:wf}
\lambda=\frac{h}{p}\simeq\frac{\hbar}{\sqrt{2mkT}} \;.
\end{equation}
becomes comparable to the average inter-particle spacing.  
In a transition to a quantum liquid state the constituent particles 
are delocalized and the system reaches a homogeneous density. 
Both the bosonic/fermionic characteristic of a many-body system and 
the inter-particle interaction determine the properties of a 
quantum liquid \cite{pines}. 

The cluster to liquid transition in atomic nuclei characterized by 
a typical inter-nucleon distance r$_0$ at saturation density, 
can be analyzed by considering three concomitant length-scale ratios. 
In addition to {$\lambda$/r$_0$} \cite{zin13}, where $\lambda$ is given 
by (\ref{eq:wf}), one can also consider b$_0$/r$_0$, with 
\begin{equation}\label{eq:b0}
b_0\hat{=}\frac{\hbar}{\sqrt{2mV'_0}}
\end{equation}
and V'$_0$ corresponds to the typical magnitude of the inter-particle 
interaction (V'$_0\simeq$100 MeV in the case of the nucleon-nucleon
interaction). This ratio is related to the 
quantality parameter $\Lambda$ introduced by B. Mottelson in Ref.~\cite{mot96}:
\begin{equation}
\label{eq:lam}
\Lambda\hat{=}2\left(\frac{b_0}{r_0}\right)^2= \frac{\hbar^2}{m r_0^2V'_0}  \;
\end{equation}
The quantality $\Lambda$ is defined as the ratio of the 
zero-point kinetic energy of the confined particle to its potential energy. 
The liquid phase corresponds to $\Lambda > 0.1$, whereas the 
crystalline solid phase is characterized by values of 
$\Lambda < 0.1$. Finally, one can also consider the localization 
parameter $\alpha$ introduced in Ref.~\cite{ebr12}, that takes 
into account the finite size of the system:
\begin{equation}\label{eq:alpha}
\alpha\hat{=}\frac{\Delta r}{r_0} \simeq \frac{\sqrt{\hbar}A^{1/6}}{(2mV_0r_0^2)^{1/4}}\;, 
\end{equation}
with $\Delta r$ denoting the spatial dispersion of the single-particle
wave function, A the number of constituents of the system and V$_0$
the depth of the confining potential. The right hand-side of this
relation corresponds to the case when the single-nucleon potential is
approximated by an isotropic harmonic oscillator potential (V$_0\simeq
75$ MeV for the nuclear mean-field \cite{ebr12}). For $\alpha > 1$ the
single-nucleon orbits are delocalized and the system is in the Fermi
liquid phase. For $\alpha \sim 1$ one finds a transition from the
quantum liquid phase to a hybrid phase of cluster states \cite{ebr13}.

The parameters defined by Eqs.~(\ref{eq:wf}), (\ref{eq:lam}) 
and (\ref{eq:alpha}) can be used to characterize quantum phases of 
matter and, in particular, nuclear matter. The formation and dissociation  
of clusters in nuclear matter as a function of density is determined by 
their binding energy due to Pauli blocking that leads to the Mott effect 
for vanishing binding \cite{rop}. The formation of nuclear clusters is 
similar to a transition from a superfluid to a Mott insulator phase in a 
gas of ultracold atoms held in a three-dimensional optical lattice potential 
\cite{gre02}. As the potential depth of the lattice is increased, a transition 
is observed from a phase in which each atom is spread out over the entire 
lattice, to the insulating phase in which atoms are localized with no 
phase coherence across the lattice. In the crust of a neutron star the
transition from the Wigner crystal phase to the quantum liquid also
proceeds through a cluster phase as a function of density \cite{lat}.  
Clustering occurs as a transition between the quantum 
liquid and solid state phases because of frustration effects, that is, 
due to the interplay between an attractive and a
repulsive interaction \cite{na07,sator}. This is the case in the
crust of neutron stars or for gelification in condensed matter.

The aim of this article is to analyze in more detail the role of
saturation for cluster formation in finite nuclear systems. Cluster
states can occur in light nuclei and, generally, in dilute nuclear
systems \cite{4,bri73,rop}. In light nuclei deformation can favor
clustering because of a local increase in density toward its
saturation value and, therefore, an increase in the binding of the
system. In Ref. \cite{ebr13} we have shown that, contrary to the case
of the crust of a neutron star, crystal-like structures cannot be
formed in a nucleus. It should be noted that another possibility for
the formation of nuclear clusters or even a crystal phase is to heat
the nucleus. Using Eq. (\ref{eq:wf}) one can estimate the temperature
at which clusters form ($\lambda \sim r_0$) to be 10 MeV. This is in
quantitative agreement with recent studies of critical temperatures
for $\alpha$-particle condensation in nuclear matter
\cite{rop98,sogo}, and shows the relevance of the parameter $\lambda /
r_0$ for cluster formation. Pauli blocking can be considered as a
leading mechanism that suppresses clusterization at higher densities,
also at finite temperatures \cite{rop09}. In the present study,
however, we focus on cluster formation in light nuclei, without
considering effects of finite temperature.

We first consider the role of deformation in the formation of clusters
and relate it to the saturation of nucleonic matter. The localization
parameter (\ref{eq:alpha}) establishes a link between clusterization
and the single-nucleon spectrum \cite{ebr12,ebr13}. The relationship
between alpha clusters and single particle states in deformed nuclei
is well known \cite{kan01,hor11,abe94}. Rae~\cite{rae89} and others
\cite{4} predicted that the degeneracy of single-nucleon states at a
given deformation could generate clusters because of levels crossing.
As already suggested by Aberg \cite{abe94,fre95}, an isolated
single-particle state of the single-particle energy spectrum in a
deformed self-conjugate $N=Z$ nucleus can correspond to an
alpha-cluster, due to both the Kramers (time-invariance) degeneracy
and the isospin symmetry: two protons and two neutrons have similar
wave functions, and the localization of these functions facilitates
the formation of alpha-clusters. 

\begin{figure}[!hbt]
      \includegraphics[width=0.5\textwidth]{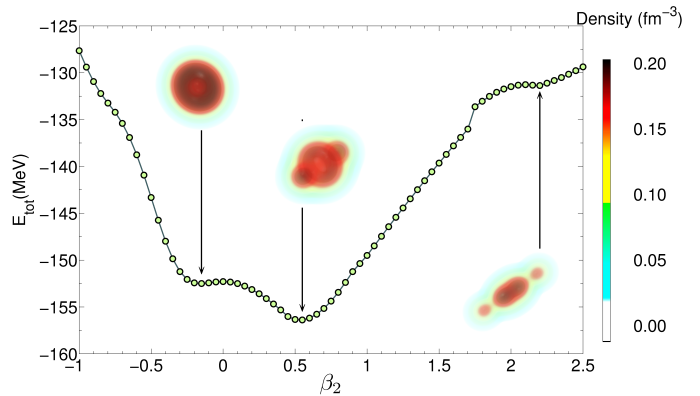}   
    \caption{(Color online) Self-consistent deformation energy curve of $^{20}$Ne as a
    function of the quadrupole deformation parameter $\beta_2$, calculated using 
    the RHB model with the DDME2 functional. The insets display the 
    corresponding three dimensional intrinsic nucleon density distributions.}
    \label{fig:meande}
  \end{figure}

\begin{figure}[!hbt]
      \includegraphics[width=0.5\textwidth]{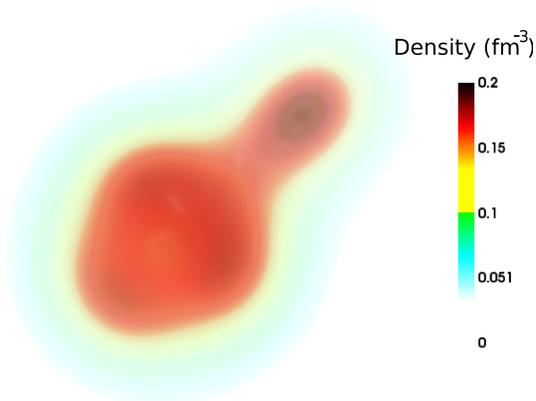}   
    \caption{(Color online) Self-consistent reflection-asymmetric 
axial intrinsic density of $^{20}$Ne, calculated with by imposing 
constraints on both the axial quadrupole and octupole deformation parameters. 
$\beta_2 = 0.55$ corresponds to the equilibrium quadrupole deformation, and 
$\beta_3 = 0.50$.}
    \label{fig:oct_quad}
  \end{figure}

To illustrate the effect of nuclear deformation 
we employ the framework of deformation-constrained self-consistent
mean-field calculations based on microscopic energy density
functionals (EDFs). This approach has recently been successfully
applied in studies of cluster phenomena in light and medium-heavy
nuclei \cite{aru05,rob11,rein,ich11,ebr12,ich12,ebr13,sch}.  It has an
advantage over dedicated cluster models in that it does not {\it a
priori} assume the existence of such structures, cluster formation is
described starting from microscopic single-nucleon degrees of freedom,
and applications are not limited only to the lightest nuclei
\cite{sch}. In fact, microscopic EDFs implicitly include many-body
correlations that enable the formation of nucleon cluster structures.
In Fig.~\ref{fig:meande} we display the binding energy of the
self-conjugate nucleus $^{20}$Ne as a function of the axial quadrupole
deformation parameter $\beta_2$. As in our previous studies of nuclear
clustering \cite{ebr12,ebr13}, the self-consistent relativistic
Hartree-Bogoliubov (RHB) model \cite{vre05} has been employed in the
calculation, based on the energy density functional DD-ME2 \cite{23}. 
The curve of the total energy as a function of quadrupole deformation
is obtained in a self-consistent mean-field calculation by imposing a
constraint on the axial quadrupole moment. The parameter $\beta_2$ is
directly proportional to the intrinsic quadrupole moment. $\beta_2 >
0$ corresponds to axial prolate shapes, whereas the shape is oblate
for $\beta_2 < 0$. The calculated equilibrium shape of $^{20}$Ne is a
prolate, axially symmetric quadrupole ellipsoid with $\beta_2 \approx
0.55$, and the characteristic observables (binding energy, charge and
matter radii) reproduce the available data within 1\%. For the
equilibrium deformation and two additional values of $\beta_2$, in the
inserts of Fig.~\ref{fig:meande} we also include the corresponding
intrinsic nucleon density distributions in the reference frame defined
by the principal axes of the nucleus.  As noted in
Refs.~\cite{ebr12,ebr13}, the equilibrium self-consistent solution
calculated with DD-ME2 yields two regions of pronounced nucleon
localization at the outer ends of the symmetry axis, and an oblate
deformed core.  Thus the intrinsic density displays a quasimolecular
$\alpha$-$^{12}$C-$\alpha$ structure. The pronounced density peaks
enhance the probability of formation of $\alpha$-clusters in excited
states close to the energy threshold for $\alpha$-particle emission
\cite{ikeda68,4,oko12,freer12}. This is clearly seen in the axial
prolate density plots for values of the deformation parameter $\beta_2
> 2$, as well as in Fig.~\ref{fig:oct_quad} where we show the
self-consistent reflection-asymmetric axial intrinsic density of
$^{20}$Ne, calculated with DD-ME2 by imposing constraints on both the
axial quadrupole and octupole deformation parameters $\beta_2$ and
$\beta_3$, respectively. For these particular values of the
deformation parameters ($\beta_2 = 0.55$ corresponds to the
equilibrium quadrupole deformation, and $\beta_3 = 0.50$), the
intrinsic density clearly presents the structure of an $^{16}$O core
plus the $\alpha$-cluster. 

%------------------------------------------------------------------------------------------------

%------------------------------------------------------------------------------------------------

The intrinsic densities shown in Figs.~\ref{fig:meande} and
\ref{fig:oct_quad} display localized lattice-like structures,
characteristic for the self-consistent mean-field approach that fixes
the center-of-mass coordinates of $\alpha$-like clusters. The 
RHB self-consistent solution contains the energy of the 
spurious center-of-mass motion of each $\alpha$-cluster that needs  
to be subtracted. A fully microscopic method for this subtraction 
has not been developed yet and, thus, a heuristic procedure was
adopted in the analysis of Ref.~\cite{sch}, that yields an extra 
binding of $\approx 7$ MeV per $\alpha$ particle.
By restoring broken
symmetries (translational, rotational, and parity in the case of
octupole deformations), and allowing for configuration mixing, one
would obtain solutions that correspond to non-localized clusters. The
concept of non-localized clustering has recently been investigated
using an angular-momentum-projected version of the 
Tohsaki-Horiuchi-Schuck-R\"opke (THSR) wave function \cite{Zhou13}. In
light nuclei at low densities $\alpha$-like clusters display a
strong tendency to condense in the same orbital with respect to their
center-of-mass motion \cite{yam12}. Deforming the nucleus by imposing
constraints on the mass multipole moments leads to excited
configurations in which the single-nucleon density is reduced along
the deformation axis with respect to the equilibrium.  Because of the
saturation property of the inter-nucleon interaction, that is, due to
the fact that the energy of nucleonic matter displays a pronounced
minimum at an equilibrium density of $\rho_{\rm eq} \approx 0.16$
fm$^{-3}$, the nucleus strengthens the binding by increasing the
density locally. For a relatively light nucleus, and especially for
self-conjugate systems, the most effective way to increase the density
locally is the formation of $\alpha$-clusters. This effect has very
recently been investigated by Girod and Schuck in Ref.~\cite{sch}. By
performing constrained HFB calculations of self-conjugate nuclei, with
a restriction to spherically symmetric configurations, they have shown
that by expanding an $n-\alpha$ nucleus the corresponding total energy
as a function of the nuclear radius goes over a maximum before
reaching the asymptotic low density limit of a gas of
$\alpha$-particles. 

\begin{figure}[!ht]
\scalebox{0.6}{\includegraphics[width=0.8\textwidth]{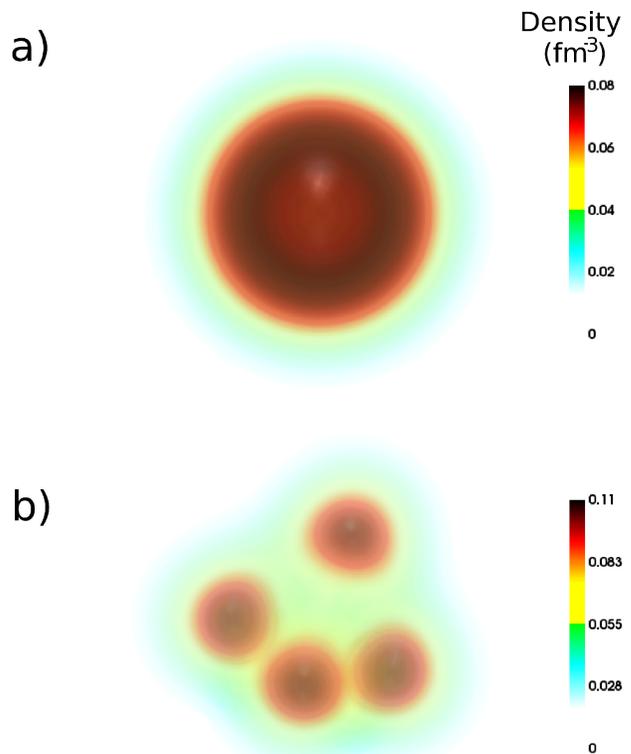}}
        \caption{(Color online) Self-consistent intrinsic nucleon density of $^{16}$O for a radius 
    constrained to 3.32 fm (a) and 3.34 fm (b).}
    \label{fig:O16_R_dens}
  \end{figure}

In Fig.~\ref{fig:O16_R_dens} we illustrate the role saturation plays
in the formation of $\alpha$-clusters in dilute nucleonic matter.  In
a self-consistent calculation similar to the one of Ref.~\cite{sch}
but using the relativistic functional DD-ME2, a constraint on the
nuclear radius is used to gradually reduce the density of $^{16}$O by
inflating the spherical nucleus. As the spherical nucleus increases in
size, the total energy of the system also increases with respect to
the equilibrium configuration. However, when the density is reduced to
$\rho / \rho_{\rm eq} \approx 1/3$, the system undergoes a Mott-like
phase transition \cite{rop,sch} to a configuration of 4
$\alpha$-particles, thus locally strengthening the binding due to the
saturation property of the inter-nucleon interaction. This transition 
occurs at a radius of $r_c = 3.33$ fm, as shown in
Fig.~\ref{fig:O16_R_dens}. The corresponding ratio of the critical
radius to the ground-state radius $r_c/ r_{g.s.} \approx 1.3$ is
smaller than the value $\approx 1.7$ obtained with the Gogny effective
interaction in Ref.~\cite{sch}. This can be explained by the fact that
single-nucleon localization in the equilibrium intrinsic density
calculated with the relativistic functional DD-ME2 is more pronounced
\cite{ebr12,ebr13} and, therefore, facilitates the formation of
$\alpha$-clusters in excited states.

Saturation therefore plays a crucial role in the emergence of clusters
in self-bound systems such as nuclei.  In a saturated system there is
a natural length scale - the equilibrium inter-particle distance,
which in nuclei is r$_0\simeq$1.2 fm. Because of this characteristic
length scale, nucleons tend to form clusters when the spatial
dispersion of the single-particle wave function is of the order of
r$_0$. Eq. (\ref{eq:alpha}) allows to show that in a large nucleus the
localization parameter $\alpha$ increases since, as it is well known, 
V$_0$ remains rather constant due to saturation. Because of the approximate
$A^{1/6}$ dependence of $\alpha$, medium-heavy and heavy nuclei will
exhibit a quantum liquid behavior, whereas cluster states can occur in
light nuclei.

By inserting Eq.~(\ref{eq:lam}) into Eq.~(\ref{eq:alpha}) one can relate the 
localization and quantality parameters:
%--------------------------------------------------------------------------
\begin{equation} \label{eq:rel}
\alpha=A^{1/6}\left(\frac{\Lambda}{2\gamma}\right)^{1/4} \;,
\end{equation}
%--------------------------------------------------------------------------
where $V_0 = \gamma {V'}_0$.  Clustering occurs for 
$\alpha\simeq$1 \cite{ebr13}, that is, when the spatial dispersion of 
the single-particle wave function is of the same order as the 
typical inter-particle distance. Inserting this last condition into 
Eq. (\ref{eq:rel}) yields an estimate for the typical
nucleon number A$_0$ at which one could expect clusters to occur:
%--------------------------------------------------------------------------
\begin{equation} \label{eq:rel3}
A_0\simeq\left(\frac{2\gamma}{\Lambda}\right)^{3/2}
\end{equation}
%--------------------------------------------------------------------------

\begin{figure}[!]
\begin{center}
\scalebox{0.35}{\includegraphics{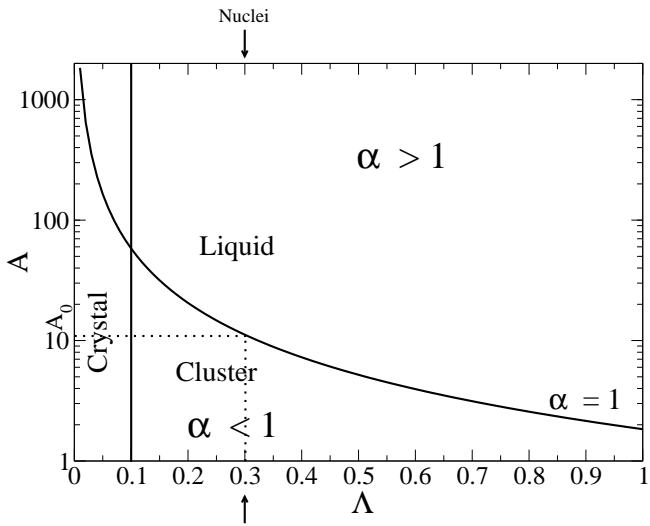}}
\caption{Phase diagram obtained from Eq. (\ref{eq:rel3}), with
$\gamma$=3/4. A$_0$ is the nucleon number for which 
clustering effects are likely to occur. $\Lambda$ is the
quantality parameter, and the arrow denotes the typical 
value obtained for nuclear matter \cite{mot96}.}
\label{fig:a0}
\end{center}
\end{figure}

For typical values of $\gamma$ in nuclei ($\gamma$ $\sim$ 3/4), and 
$\Lambda$ in nuclear matter \cite{mot96}, one finds A$_0\simeq$10 in
agreement with the mass region where cluster effects are
observed \cite{4}.  Figure \ref{fig:a0} displays the corresponding
phases in finite saturated systems, with the curve that separates the 
liquid and cluster phases determined by Eq. (\ref{eq:rel3}). This 
analysis supports the interpretation of clusters as a hybrid phase: 
for a system to behave as a liquid, $\Lambda$ should be greater than
0.1 but also $A > A_0$ ($\alpha > 1$), and this underlines the
importance of finite size effects in the formation of clusters. The
cluster phase corresponds to $\Lambda >$ 0.1 and $\alpha <$ 1. In
other words, the relation between $\alpha$ and $\Lambda$ shows that in
the cluster phase the confining potential attenuates the
single-nucleon quantum-liquid delocalization generated by the
inter-nucleon interaction.

In summary, we have analyzed the role of saturation in the mechanism
of cluster formation in finite nuclei and in dilute nuclear matter.
The localization parameter describes how saturation allows for cluster
and quantum liquid phases of nuclei by relating them with the
single-particle behavior through the depth of the confining potential.
In deformed light nuclei the formation of clusters is favored because
it locally enhances the nucleonic density toward its saturation value,
thus increasing the binding of the system. In those nuclei the
confining potential weakens the quantum-liquid delocalisation induced
by the inter-nucleon interaction. In general, when the density of
nucleonic matter is reduced below its equilibrium values, saturation
causes a Mott-like transition to a hybrid phase composed of clusters
of $\alpha$-particles.

\bigskip
This work was supported by the Institut Universitaire de France. The
authors thank Peter Schuck for reading the manuscript and many valuable 
discussions.

\end{document}